\documentclass[a4paper,12pt]{article}
\usepackage{amssymb,amsmath,amsfonts}
\usepackage{fancyhdr}
\usepackage{a4}
\usepackage{parskip}
\usepackage[all,2cell]{xy}\UseAllTwocells

\newcommand{\sect}[1]{\setcounter{equation}{0}\section{#1}}

\def\ep{\varepsilon}

\def\axs{AdS_5\times S^5}
\newcommand{\eq}[1]{\begin{equation} #1 \end{equation}}
\newcommand{\al}[1]{\begin{align} #1 \end{align}}

\begin{document}
\begin{titlepage}
\markright{\bf TUW--10--09}
\title{On the pulsating strings in $AdS_5\times T^{1,1}$}

\author{D. Arnaudov${}^{\star}$, H. Dimov${}^{\star}$ and R.~C.~Rashkov${}^{\dagger,\star}$\thanks{e-mail:
rash@hep.itp.tuwien.ac.at.}
\ \\ \ \\
${}^{\star}$  Department of Physics, Sofia
University,\\
5 J. Bourchier Blvd, 1164 Sofia, Bulgaria
\ \\ \ \\
${}^{\dagger}$ Institute for Theoretical Physics, \\ Vienna
University of Technology,\\
Wiedner Hauptstr. 8-10, 1040 Vienna, Austria
}
\date{}
\end{titlepage}
\maketitle
\thispagestyle{fancy}

\begin{abstract}
We study the class of pulsating strings in $AdS_5\times T^{1,1}$. Using a generalized ansatz for pulsating string configurations we find new solutions of this class. Further we semiclassically quantize the theory and obtain the first correction to the energy. The latter, due to AdS/CFT correspondence, is supposed to give the anomalous dimensions of operators in the dual ${\cal N}=1$ superconformal gauge field theory.
\end{abstract}

\sect{Introduction}

The attempt to establish a correspondence between the large $N$ limit of gauge theories and string theory has more than 30 years history and
over the years it showed different faces. Recently an explicit realization of this correspondence was provided by the Maldacena conjecture about AdS/CFT correspondence \cite{Maldacena}. The convincing results from the duality between type IIB string theory on $\axs$ and ${\cal N}=4$ SYM theory~\cite{Maldacena,GKP,Witten} made this subject a major research area, and many fascinating new features have been established.

After the impressive achievements in the most supersymmetric example of AdS/CFT correspondence, namely $\axs$, it is important to extend the considerations to less supersymmetric gauge theories, moreover that the latter are more interesting from physical point of view. There are several ways to find a theory with less supersymmetry. The experience from AdS/CFT correspondence suggests that one of them is to take a stack of $N$ D3 branes and place them not in flat space, but at the apex of a conifold~\cite{Klebanov:1998hh}. This model possesses a lot of interesting features and allows to build gauge theory operators of great physical importance, although there are hints that superstring theory on the resulting ten-dimensional spacetime, which takes form as the direct product $AdS_5\times T^{1,1}$, is not integrable \cite{Basu:2011}. Since \cite{Klebanov:1998hh}, infinite families of five-dimensional Sasaki-Einstein spaces, complementing the AdS space, have been constructed \cite{Gauntlett:2004yd,Cvetic:2005ft}, and also their gauge theory duals were identified \cite{Benvenuti:2004dy}--\cite{Butti:2005sw}. Further developments can be traced in \cite{Gauntlett:2004yd}--\cite{Giataganas:2009}.

Semiclassical strings have played, and still play, an important role in studying various aspects of AdS$_5$/SYM$_4$ correspondence \cite{Bena:2003wd}--\cite{Lee:2008sk}. The development in this subject gives strong hints about how the new emergent duality can be investigated. An important class of semiclassical string solutions is the class of pulsating strings introduced first in~\cite{Minahan:2002rc}, and generalized further in \cite{Engquist:2003rn,Dimov:2004xi,Smedback:2004}. In the case of $AdS_5\times T^{1,1}$ background the pulsating strings are also expected to play an essential role, but thorough analysis and semiclassical quantization are still missing. The purpose of this paper is to analyze and semiclassically quantize the class of pulsating strings on the $T^{1,1}$ part of the $AdS_5\times T^{1,1}$ background. The first correction to the energy, which according to the AdS/CFT conjecture gives the anomalous dimensions of gauge theory operators, will be the main subject of our considerations.

The paper is organized as follows. In the introduction we provide some details about the dual gauge theory. In the next section we present the pulsating
strings and their semiclassical quantization for the case of $AdS_5\times T^{1,1}$ background, restricting the string dynamics to the $T^{1,1}$ part. The third section is devoted to the derivation of the correction to the energy. First we find the wave function associated with the Laplace-Beltrami operator on $T^{1,1}$, and then we compute the leading correction to the energy. We conclude with a brief discussion on the results.

\paragraph{Dual field theory}

The respective ${\cal N}=1$ superconformal gauge theory dual to string theory in $AdS_5\times T^{1,1}$ is known as the Klebanov-Witten theory and was originally described in \cite{Klebanov:1998hh}. The theory has flavor symmetry $SU(2)\times SU(2)$. The elementary degrees of freedom are denoted by the fields $A$ and $B$, each a doublet of the factor $SU(2)$ groups and with conformal anomalous dimensions $\Delta_{A,B}=3/4$. The gauge group is $SU(N)\times SU(N)$, and the two chiral multiplets $A$ and $B$ are correspondingly in the $(N,\bar{N})$ and $(\bar{N},N)$ representations. The superpotential is
\eq{
W=\frac{\lambda}{2}\,\ep^{ij}\ep^{kl}\,{\rm Tr}[A_iB_kA_jB_l]\,,
}
where $i=1,2$. The chiral operators analogue of the $(X,Y,Z)$ operators in ${\cal N}=4$ SYM are given by ${\rm Tr}(AB)^k$ with R-charge $k$ and in the
$(\frac{k}{2},\frac{k}{2})$ representation of the flavor group $SU(2)\times SU(2)$.

\sect{Pulsating strings in $AdS_5\times T^{1,1}$}

In this section we consider a circular pulsating string expanding and contracting only on the $T^{1,1}$ part of $AdS_5\times T^{1,1}$ ($\mathbb{R}\times T^{1,1}$). Then, the relevant metric we will work with is given by
\begin{equation}
ds^2_{t\times T^{1,1}}=R^2\!\left(-dt^2+ds^2_{T^{1,1}}\right),
\end{equation}
where the metric of $T^{1,1}$ is
\begin{equation}
ds^2_{T^{1,1}}=\frac{b}{4}\!\left[\sum\limits_{i=1}^2\left(d\theta_i^2+\sin^2\theta_id\phi_i^2\right)
+b\!\left(d\psi+\sum\limits_{i=1}^2\cos\theta_id\phi_i\right)^{\!\!2}\,\right]\!,
\label{metric_T11}
\end{equation}
where $0\leq\psi<4\pi,\ 0\leq\theta_i\leq\pi,\ 0\leq\phi_i<2\pi$, and $b=2/3$ (for $b=1$ we recover $S^5$). Having in mind the explicit form of the $T^{1,1}$ metric \eqref{metric_T11}, one can write it as
\begin{equation}
ds^2_{T^{1,1}}=G_{ij}dx^idx^j+\hat{G}_{pq}dy^pdy^q,
\end{equation}
where $G_{ij}$ is defined by
\begin{equation}
G_{ij}={\rm diag}\!\left(\frac{b}{4},\,\frac{b}{4}\right)\!,\ i,j=1,2\,,\quad x^1=\theta_1\,,\quad x^2=\theta_2\,,
\end{equation}
while $\hat{G}_{pq}=\hat{G}_{pq}(\theta_1,\theta_2)$ is the remaining part of the metric, associated with $\phi_1,\phi_2,\psi$ coordinates, which are denoted here as $p,q=1,2,3,\ y^1=\phi_1,\ y^2=\phi_2,\ y^3=\psi$
\begin{equation}
(\hat{G}_{pq})=\dfrac{b}{4}\!\left(
\begin{array}{ccc}
b\cos^2\theta_1+\sin^2\theta_1 & b\cos\theta_1\cos\theta_2 & -b\cos\theta_1\\
b\cos\theta_1\cos\theta_2 & b\cos^2\theta_2+\sin^2\theta_2 & -b\cos\theta_2\\
-b\cos\theta_1 & -b\cos\theta_2 & b
\end{array}
\right)\!.
\end{equation}
The residual worldsheet symmetry allows to identify $t$ with $\tau$, and to obtain a classical pulsating string solution we use the following ansatz
\al{
x^1&=x^1(\tau)=\theta_1(\tau)\,,\qquad\qquad\!x^2=x^2(\tau)=\theta_2(\tau)\,,\\
y^1&=\phi_1=m_1\sigma+h^1(\tau)\,,\qquad y^2=\phi_2=m_2\sigma+h^2(\tau)\,,\qquad y^3=\psi=m_3\sigma+h^3(\tau)\,.
}
We are interested in the induced worldsheet metric, which in our case has the form
\begin{equation}
ds^2_{\rm ws}=R^2\!\left[\left(-1+G_{ij}\dot{x}^i\dot{x}^j+\hat{G}_{pq}\dot{h}^p\dot{h}^q\right)d\tau^2+(\hat{G}_{pq}m_p m_q)d\sigma^2+
2(\hat{G}_{pq}m_p\dot{h}^q)d\tau d\sigma\right]\!.\!\!\!
\end{equation}
The Nambu-Goto action
\begin{equation}
S_{\rm NG}=-T\int d\tau d\sigma\,\sqrt{-{\rm det}(G_{\mu\nu}\,\partial_\alpha X^{\mu}\,\partial_\beta X^{\nu})}
\end{equation}
in this ansatz then reduces to the expression
\begin{equation}
S_{\rm NG}=-TR^2\int d\tau d\sigma\,\sqrt{\left(1-G_{ij}\dot{x}^i\dot{x}^j-\hat{G}_{pq}\dot{h}^p\dot{h}^q\right)(\hat{G}_{pq}m_p m_q)+
(\hat{G}_{pq}m_p\dot{h}^q)^2}\,,
\end{equation}
where $TR^2=\sqrt{\lambda}$. For our considerations it is useful to pass to Hamiltonian formulation. For this purpose, we have to find first the canonical momenta of our dynamical system. Straightforward calculations yield
\al{
\Pi_i&=\sqrt{\lambda}\,\dfrac{(\hat{G}_{pq}m_pm_q)G_{ij}\dot{x}^j}{\sqrt{\left(1-G_{ij}\dot{x}^i\dot{x}^j-\hat{G}_{pq}\dot{h}^p\dot{h}^q\right)(\hat{G}_{pq}m_pm_q)+
(\hat{G}_{pq}m_p\dot{h}^q)^2}}\,,\ i=1,2\,,\\
\hat{\Pi}_p&=\sqrt{\lambda}\,\dfrac{(\hat{G}_{pq}m_pm_q)\hat{G}_{pq}\dot{h}^q-(\hat{G}_{pq}m_p\dot{h}^q)\hat{G}_{pq}m_q}
{\sqrt{\left(1-G_{ij}\dot{x}^i\dot{x}^j-\hat{G}_{pq}\dot{h}^p\dot{h}^q\right)(\hat{G}_{pq}m_pm_q)+(\hat{G}_{pq}m_p\dot{h}^q)^2}}\,,\ p=1,2,3\,,
}
which also implies the constraint
\begin{equation}
m_p\hat{\Pi}_p=0\,.
\label{impconstr}
\end{equation}
Solving for the derivatives in terms of the canonical momenta and substituting back into the Legendre transform of the Lagrangian, we find the Hamiltonian
\begin{equation}
H^2=G^{ij}\Pi_i\Pi_j+\hat{G}^{pq}\hat{\Pi}_p\hat{\Pi}_q+\lambda\,(\hat{G}_{pq}m_pm_q)\,.
\end{equation}
The interpretation of this relation is as in the case of $\axs$ \cite{Dimov:2004xi}. Namely, the first two terms represent kinetic energy, while the last one is considered as a potential $V$, which in our case has the form
\begin{equation}
V(\theta_1,\theta_2)=\lambda\,\hat{G}_{pq}(\theta_1,\theta_2)m_pm_q\,.
\end{equation}
The approximation where our considerations are valid assumes high energies, which suggests that one can think of this potential term as a perturbation. For later use we write down the explicit form of the potential
\al{\nonumber
V(\theta_1,\theta_2)&=\lambda\,\dfrac{b}{4}\left[\left(b\cos^2\theta_1+\sin^2\theta_1\right)\!m_1^2+\left(b\cos^2\theta_2+\sin^2\theta_2\right)\!m_2^2+bm_3^2\right.\\
&\left.+\,2b\cos\theta_1\cos\theta_2m_1m_2-2b\cos\theta_1m_1m_3-2b\cos\theta_2m_2m_3\right],
\label{pot1}
}
or
\begin{equation}
V(\theta_1,\theta_2)=\lambda\,\dfrac{b}{4}\!\left[\sum\limits_{i=1}^2m_i^2\sin^2\theta_i+b\!\left(m_3-\sum\limits_{i=1}^2m_i\cos\theta_i\right)^{\!\!2}\,\right]\!.
\label{pot2}
\end{equation}
The above perturbation to the free action will produce the correction to the energy and therefore the anomalous dimension. In order to calculate the correction to the energy as a perturbation due to the above potential, however, we need the normalized wave function associated with $T^{1,1}$ space. All these issues are subject to the next section.

\sect{Semiclassical correction to the energy}

In this section we will compute the semiclassical correction to the energy of the pulsating string on $T^{1,1}$. As we discussed in the previous section, the Hamiltonian of the pulsating string is interpreted as a dynamical system with high energy described by free theory Schr\"{o}dinger equation and perturbed by the potential \eqref{pot1}.

First of all, we have to find the wave function associated with the Laplace-Beltrami operator on $T^{1,1}$ and then to obtain the correction to the energy due to the induced potential.

\subsection{Laplace-Beltrami operator and wave function}

The line element of $T^{1,1}$ in global coordinates is explicitly given by \eqref{metric_T11}
\begin{equation}
ds^2_{T^{1,1}}=\frac{b}{4}\!\left[\sum\limits_{i=1}^2\left(d\theta_i^2+\sin^2\theta_id\phi_i^2\right)
+b\!\left(d\psi+\sum\limits_{i=1}^2\cos\theta_id\phi_i\right)^{\!\!2}\,\right]\!.
\end{equation}

\paragraph{Laplace-Beltrami operator on $T^{1,1}$}

Using the standard definition of the Laplace-Beltrami operator in global coordinates we find (see section 3 of \cite{Gubser:1999} for the general case of $T^{p,q}$)
\al{
&\bigtriangleup_{T^{1,1}}=\dfrac{4}{b^2}\!\left[\dfrac{b}{\sin\theta_1}\dfrac{\partial}{\partial\theta_1}\!\left(\!\sin\theta_1\dfrac{\partial}{\partial\theta_1}\!\right)\!+
\dfrac{b}{\sin^2\theta_1}\dfrac{\partial^2}{\partial\phi_1^2}+\dfrac{2b\cos\theta_1}{\sin^2\theta_1}\dfrac{\partial^2}{\partial\phi_1\partial\psi}\right.\\
&\left.+\,\dfrac{b}{\sin\theta_2}\dfrac{\partial}{\partial\theta_2}\!\left(\!\sin\theta_2\dfrac{\partial}{\partial\theta_2}\!\right)\!+
\dfrac{b}{\sin^2\theta_2}\dfrac{\partial^2}{\partial\phi_2^2}+\dfrac{2b\cos\theta_2}{\sin^2\theta_2}\dfrac{\partial^2}{\partial\phi_2\partial\psi}+
\!\left(\!1+\dfrac{b\cos\theta_1}{\sin^2\theta_1}+\dfrac{b\cos\theta_2}{\sin^2\theta_2}\!\right)\!\dfrac{\partial^2}{\partial\psi^2}\right]\!,
\nonumber
}
or
\al{\label{laplas}
\bigtriangleup_{T^{1,1}}&=\dfrac{4}{b^2}\!\left\lbrace
b\!\left[\dfrac{1}{\sin\theta_1}\dfrac{\partial}{\partial\theta_1}\!\left(\sin\theta_1\dfrac{\partial}{\partial\theta_1}\right)\!+
\dfrac{1}{\sin^2\theta_1}\!\left(\dfrac{\partial}{\partial\phi_1}+\cos\theta_1\dfrac{\partial}{\partial\psi}\right)^{\!\!2}\,\right]\right.\\
&\left.+\,b\!\left[\dfrac{1}{\sin\theta_2}\dfrac{\partial}{\partial\theta_2}\!\left(\sin\theta_2\dfrac{\partial}{\partial\theta_2}\right)\!+
\dfrac{1}{\sin^2\theta_2}\!\left(\dfrac{\partial}{\partial\phi_2}+\cos\theta_2\dfrac{\partial}{\partial\psi}\right)^{\!\!2}\,\right]\!+
\dfrac{\partial^2}{\partial\psi^2}\right\rbrace.
\nonumber
}
The full measure on $T^{1,1}$ is
\begin{equation}
d\Omega(\theta_1,\theta_2)=\sqrt{{\rm det}(G_{\mu\nu})}\,d\theta_1d\theta_2=2\!\left(\frac{b}{4}\right)^3\sin\theta_1\sin\theta_2\,d\theta_1d\theta_2\,.
\label{measure1}
\end{equation}

\paragraph{Wave function}

The Schr\"{o}dinger equation for the wave function is
\begin{equation}
\bigtriangleup_{T^{1,1}}\,\Psi(\theta_1,\theta_2,\phi_1,\phi_2,\psi)=-E^2\,\Psi(\theta_1,\theta_2,\phi_1,\phi_2,\psi)\,.
\label{schro}
\end{equation}
To separate the variables, we define $\Psi$ as
\begin{equation}
\Psi(\theta_1,\theta_2,\phi_1,\phi_2,\psi)=f_1(\theta_1)\,f_2(\theta_2)\,f(\phi_1,\phi_2,\psi)\,,
\label{wave}
\end{equation}
where
\begin{equation}
f(\phi_1,\phi_2,\psi)=\exp(il_1\phi_1)\exp(il_2\phi_2)\exp(il_3\psi)\,,\qquad l_1,\,l_2,\,l_3\in\mathbb{Z}\,.
\end{equation}
With this choice we can solve for the eigenfunctions, replacing the derivatives along Killing directions $(\partial_{\phi_1},\partial_{\phi_2},\partial_\psi)$ by $(il_1,il_2,il_3)$ correspondingly. Note that \eqref{impconstr} implies the following relation among the parameters $l_1,l_2,l_3$
\begin{equation}
m_1l_1+m_2l_2+m_3l_3=0\,.
\end{equation}
Substituting \eqref{wave} in \eqref{schro}, together with \eqref{laplas}, we arrive at
\begin{equation}
E^2=\dfrac{4}{b^2}(bE_1^2+bE_2^2+l_3^2)\,,
\label{energy^2}
\end{equation}
where $E_1$ and $E_2$ are determined by the ordinary differential equations
\begin{equation}
\left[\dfrac{1}{\sin\theta_i}\dfrac{d}{d\theta_i}\!\left(\sin\theta_i\dfrac{d}{d\theta_i}\right)\!-
\dfrac{1}{\sin^2\theta_i}\left(l_i+\cos\theta_i l_3\right)^2\right]\!f_i(\theta_i)=-E_i^2\,f_i(\theta_i)\,,\quad i=1,2\,.
\end{equation}
It is convenient to define new variables $z_i=\cos\theta_i$. Then the equations can be written as
\begin{equation}
\left((1-z_i^2)\,\dfrac{d^2}{dz_i^2}-2z_i\,\dfrac{d}{dz_i}-\dfrac{1}{1-z_i^2}\,(l_i+z_i\,l_3)^2+E_i^2\right)f_i(z_i)=0\,.
\end{equation}
The solutions to these equations are
\begin{multline}
f_i(z_i)=(1-z_i)^{\frac{\mid l_i+l_3\mid}{2}}\,(1+z_i)^{\frac{\mid l_i-l_3\mid}{2}}\ {_2F_1}\!\left[\dfrac{1}{2}\!\left(\mid\!l_i+l_3\!\mid+\mid\!l_i-l_3\!\mid+\,1-\sqrt{1+4(l_3^2+E_i^2)}\right)\!,\right.\\
\left.\dfrac{1}{2}\!\left(\mid\!l_i+l_3\!\mid+\mid\!l_i-l_3\!\mid+\,1+\sqrt{1+4(l_3^2+E_i^2)}\right)\!;\,1\,+\mid\!l_i-l_3\!\mid\,;\,\dfrac{1+z_i}{2}\right]\!.
\end{multline}
In addition, we have to ensure that the solutions $f_i(\theta_i)$ are square integrable with respect to the measure for $\theta_i$, which leads to the following restrictions on the parameters
\begin{equation}
\sqrt{1+4(l_3^2+E_i^2)}\ -\mid\!l_i+l_3\!\mid-\mid\!l_i-l_3\!\mid-\,1=\,2n_i\,,\qquad n_i\in\mathbb{N}\,.
\label{enpar}
\end{equation}
From \eqref{energy^2} and \eqref{enpar} follows that the squares of the bare dimensions of dual operators are
\begin{equation}
\Delta^2=E^2=\frac{4}{b^2}\!\left(\frac{b}{4}\sum_{i=1}^2(2n_i+\mid\!l_i+l_3\!\mid+\mid\!l_i-l_3\!\mid+\,1)^2+(1-2b)l_3^2-\frac{b}{2}\right)\!.
\end{equation}
Introducing new parameters $\alpha_i\equiv\ \mid\!l_i-l_3\!\mid$ and $\beta_i\equiv\ \mid\!l_i+l_3\!\mid$, the solution can be written in terms of Jacobi polynomials
\begin{equation}
f_i(z_i)=(1-z_i)^{\alpha_i/2}\,(1+z_i)^{\beta_i/2}\,\dfrac{n_i!\,\Gamma(\alpha_i+1)}{\Gamma(\alpha_i+1+n_i)}P^{(\alpha_i,\beta_i)}_{n_i}(z_i)\,.
\label{solution}
\end{equation}
The normalized wave functions look like
\al{\nonumber
\Psi_{n_i}^{\alpha_i,\beta_i}(z_i)&=\sqrt{\frac{(\alpha_i+\beta_i+1+2n_i)\,n_i!\,\Gamma(\alpha_i+\beta_i+1+n_i)}{2^{\alpha_i+\beta_i+1}\,
\Gamma(\alpha_i+1+n_i)\,\Gamma(\beta_i+1+n_i)}}\\
&\times(1-z_i)^{\alpha_i/2}\,(1+z_i)^{\beta_i/2}P^{(\alpha_i,\beta_i)}_{n_i}(z_i)\,,\qquad i=1,2\,.
\label{wfun}
}

\subsection{Leading correction to the energy}

It is convenient to write the potential \eqref{pot1} in the following form
\al{\nonumber
V(\theta_1,\theta_2)&=\lambda\,\dfrac{b}{4}\!\left[b\sum\limits_{i=1}^3m_i^2+(1-b)\sum\limits_{i=1}^2m_i^2\sin^2\theta_i\right.\\
&\left.+\,2bm_1m_2\cos\theta_1\cos\theta_2-2bm_3\sum\limits_{i=1}^2m_i\cos\theta_i\right]\!.
\label{pot3}
}
In terms of the new variables $z_i$ the potential and the measure \eqref{measure1} become
\al{
V(z_1,z_2)&=\lambda\,\dfrac{b}{4}\!\left[b\sum\limits_{i=1}^3m_i^2+(1-b)\sum\limits_{i=1}^2m_i^2(1-z_i^2)+2b\!\left(m_1m_2z_1z_2-m_3\sum\limits_{i=1}^2m_i z_i\right)\!\right]\!,\label{pot4}\\
d\Omega(z_1,z_2)&=2\!\left(\dfrac{b}{4}\right)^3dz_1\,dz_2\,,\quad-1\leq z_1,z_2\leq1\,.
\nonumber
}
The first correction to the energy is given then by the expression
\begin{equation}
\delta E^2=\int\limits_{-1}^1\int\limits_{-1}^1d\Omega(z_1,z_2)\,V(z_1,z_2) \left[\Psi_{n_1}^{\alpha_1,\beta_1}(z_1)\right]^2\left[\Psi_{n_2}^{\alpha_2,\beta_2}(z_2)\right]^2\!.
\label{delta-E}
\end{equation}
The explicit form of the correction to the energy is obtained by plugging the various wave functions and the potential in \eqref{delta-E}
\al{
&\delta E^2=\lambda\!\left(\dfrac{b}{4}\right)^4\!\left(b\sum\limits_{i=1}^3m_i^2+(1-b)\sum\limits_{i=1}^2m_i^2
\int\limits_{-1}^1dz_i\,(1-z_i^2)\left[\Psi_{n_i}^{\alpha_i,\beta_i}(z_i)\right]^2+2bm_1m_2\right.\\
&\left.\times\int\limits_{-1}^1dz_1\,z_1\left[\Psi_{n_1}^{\alpha_1,\beta_1}(z_1)\right]^2\int\limits_{-1}^1dz_2\,z_2\left[\Psi_{n_2}^{\alpha_2,\beta_2}(z_2)\right]^2-
2bm_3\sum\limits_{i=1}^2m_i\int\limits_{-1}^1dz_i\,z_i\left[\Psi_{n_i}^{\alpha_i,\beta_i}(z_i)\right]^2\right)\!.
\nonumber
}
In short notations it looks like
\begin{equation}
\delta E^2=\lambda\!\left(\dfrac{b}{4}\right)^4\left[b\sum\limits_{i=1}^3m_i^2+(1-b)\sum\limits_{i=1}^2m_i^2I_1^i+2bm_1m_2I_2^1I_2^2-
2bm_3\sum\limits_{i=1}^2m_iI_2^i\right]\!,
\label{corr-energy}
\end{equation}
where the integrals $I_1^i$ and $I_2^i$ are explicitly calculated as follows
\al{
I_1^i&=\int\limits_{-1}^1dz_i\,(1-z_i^2)\left[\Psi_{n_i}^{\alpha_i,\beta_i}(z_i)\right]^2\\ \nonumber
&=\frac{(n_i+\alpha_i+\beta_i+1)(n_i+\alpha_i+\beta_i+2)(n_i+\alpha_i+1)(n_i+\beta_i+1)}{(2n_i+\alpha_i+\beta_i+1)
(2n_i+\alpha_i+\beta_i+2)^2(2n_i+\alpha_i+\beta_i+3)}\\ \nonumber
&+\dfrac{n_i(n_i+\alpha_i+\beta_i+1)}{(2n_i+\alpha_i+\beta_i+1)^2}\!\left( \frac{n_i+\alpha_i+1}{2n_i+\alpha_i+\beta_i+2}-\frac{n_i+\beta_i}{2n_i+\alpha_i+\beta_i}\right)^2\\
&+\frac{n_i(n_i-1)(n_i+\alpha_i)(n_i+\beta_i)}{(2n_i+\alpha_i+\beta_i+1)(2n_i+\alpha_i+\beta_i)^2(2n_i+\alpha_i+\beta_i-1)}\,,\nonumber\\
\nonumber\\
I_2^i&=\int\limits_{-1}^1dz_i\,z_i\left[\Psi_{n_i}^{\alpha_i,\beta_i}(z_i)\right]^2\\
&=\frac{2(n_i+\beta_i)(n_i+\alpha_i+\beta_i)}{(2n_i+\alpha_i+\beta_i+1)(2n_i+\alpha_i+\beta_i)}+
\frac{2(n_i+1)(n_i+\alpha_i+1)}{(2n_i+\alpha_i+\beta_i+1)(2n_i+\alpha_i+\beta_i+2)}\,.
\nonumber
}
The expression for the correction to the energy looks very complicated. Therefore, we use the fact that the approximation we work in is for large quantum numbers, say $n_{1,2}\gg\alpha_{1,2} (\beta_{1,2})\gg1$. Within this approximation the integrals behave like
\al{
I_1^i&=\frac{1}{8}+\frac{1}{32}\!\left(2\alpha_i^2+2\beta_i^2-1\right)\!\dfrac{1}{n_i^2}+{\rm O}\!\left(\frac{1}{n_i^3}\right)\!,\\
I_2^i&=1+\frac{1}{4}\!\left(\beta_i^2-\alpha_i^2\right)\!\dfrac{1}{n_i^2}+{\rm O}\!\left(\frac{1}{n_i^3}\right)\!.
}
Since $\alpha_i\equiv\ \mid\!l_i-l_3\!\mid$ and $\beta_i\equiv\ \mid\!l_i+l_3\!\mid$, the above integrals look like
\al{
I_1^i&=\frac{1}{8}+\frac{1}{8}\!\left(l_i^2+l_3^2-\frac{1}{4}\right)\!\dfrac{1}{n_i^2}+{\rm O}\!\left(\frac{1}{n_i^3}\right)\!,\\
I_2^i&=1-\dfrac{l_il_3}{n_i^2}+{\rm O}\!\left(\frac{1}{n_i^3}\right)\!.
}
Ignoring the terms of higher order we obtain
\al{\nonumber
\delta E^2&\approx\lambda\!\left(\dfrac{b}{4}\right)^4\Bigg[b\!\left(m_3-\sum\limits_{i=1}^2m_i\right)^2+\dfrac{1-b}{8}\sum\limits_{i=1}^2m_i^2\!\left(1-
\dfrac{1}{4n_i^2}+\dfrac{l_i^2+l_3^2}{n_i^2}\right)\\
&+2b\sum\limits_{i=1}^2\,(m_im_3-m_1m_2)\,\dfrac{l_il_3}{n_i^2}\Bigg].
}

\sect{Conclusion}

Our study is motivated by the recently suggested duality between string theory in $AdS_5\times T^{1,1}$ and ${\cal N}=1$ superconformal field theory. The results obtained so far provide important understanding of string/gauge theory dualities, particularly in the region of strong coupling \cite{Klebanov:1998hh}--\cite{Giataganas:2009dr}. The purpose of this paper is to investigate the pulsating string solutions in $AdS_5\times T^{1,1}$
background. The class of pulsating strings has been used to study the AdS/CFT correspondence in the case of $\axs$ \cite{Minahan:2002rc,Dimov:2004xi,Smedback:2004}, and the leading correction to the string energy has been associated with anomalous dimensions of certain
operators in the dual gauge theory.

Here we consider a generalized string ansatz for a pulsating string in the $T^{1,1}$ part of the geometry. Next we derive the correction to the classical energy. From AdS/CFT point of view the correction gives the anomalous dimensions of operators in SYM theory and therefore it is of primary interest. For this purpose, we consider the Nambu-Goto action and find the Hamiltonian. After that we quantize the resulting theory semiclassically and obtain the correction to the energy. Since we consider a highly excited system, the kinetic term is dominating. This means that we effectively perform summation over all classical solutions (not only those that have been explicitly found), while the effective potential term serves as a small perturbation. The obtained correction to the classical energy looks complicated, but in a certain limit one can find a relatively simple expression. To identify the contributions of the different terms, it is instructive to look at the solutions for the $S^5$ case. Since they correspond to a subsector well known from $\axs$ considerations, one can identify the origin of the various contributions. One can see that the correction to the energy in $T^{1,1}$ has analogous structure to the case of pulsating strings in $S^5$, for example. The mixing between quantum numbers of different isometry directions shows up in an analogous, but slightly more complicated, way. This can be seen using the result from the $S^3$ subsector and its embedding in $T^{1,1}$.

As a final comment, we note that in order to complete the study from AdS/CFT point of view, it is of great importance to perform an analysis comparing our result to that on the SYM side. We leave this problem for future research.

\section*{Acknowledgments}
This work was supported in part by the Austrian Research Funds FWF P22000 and I192, NSFB VU-F-201/06 and DO 02-257.


\end{document}